\documentclass[conference]{IEEEtran}
\IEEEoverridecommandlockouts
\usepackage{cite}
\usepackage{amsmath,amssymb,amsfonts}
\usepackage{algorithmic}
\usepackage{graphicx}
\usepackage{textcomp}
\usepackage{xcolor}
\usepackage{nicefrac}
\usepackage{graphics}
\usepackage{float}
\usepackage{graphicx}
\usepackage{amsmath}
\usepackage{cite}
\usepackage{color}
\usepackage{dsfont}
\usepackage{bm}
\usepackage{upgreek}
\usepackage{arydshln}
\usepackage{enumerate}
\usepackage{amsthm}
\usepackage{graphicx}
\usepackage{threeparttable}
\usepackage{caption}
\usepackage{multirow}
\usepackage{color}
\usepackage{xfrac}
\usepackage{array}
\usepackage{nomencl}
\usepackage{hyphenat}
\usepackage{cite}
\usepackage{amsmath,amssymb,amsfonts}
\usepackage{algorithmic}
\usepackage{graphicx}
\usepackage{textcomp}
\usepackage{xcolor}
\def\BibTeX{{\rm B\kern-.05em{\sc i\kern-.025em b}\kern-.08em
    T\kern-.1667em\lower.7ex\hbox{E}\kern-.125emX}}

\newcommand{\be}{\begin{equation}}
\newcommand{\ee}{\end{equation}}

\newcommand{\norm}[1]{ || #1 ||}
\newcommand{\trace}[1]{ \mathrm{tr}( #1 )}
\newcommand{\mb}[1]{\mathbf{#1}}
\newcommand{\bs}[1]{\boldsymbol{#1}}

\newcommand{\virg}[1]{\textquotedblleft#1\textquotedblright}

\newcommand{\pV}{\widehat{\mb{V}}^\star_1}
\newcommand{\VR}{\widehat{\mb{V}}_{1,R}}

\newcommand{\cvec}[1]{ \mathrm{vec}\left(  #1 \right) }

\newcommand{\ovec}[1]{\underline{\mathrm{vec}}(#1)}

\newcommand{\tonde}[1]{\left( #1 \right)  }
\newcommand{\quadre}[1]{\left[  #1 \right]  }
\newcommand{\graffe}[1]{\left\lbrace   #1 \right\rbrace   }

\begin{document}

\title{Properties of a new $R$-estimator of shape matrices\\
\thanks{The work of S. Fortunati, A. Renaux and F. Pascal has been partially supported by DGA under grant ANR-17-ASTR-0015.}
}

\author{\IEEEauthorblockN{Stefano~Fortunati, Alexandre~Renaux, Fr\'{e}d\'{e}ric~Pascal}
\IEEEauthorblockA{\textit{Universit\'{e} Paris-Saclay, CNRS, CentraleSupel\'{e}c, Laboratoire des signaux et syst\`{e}mes,} \\
91190, Gif-sur-Yvette, France. \\
e-mails: \{stefano.fortunati, alexandre.renaux, frederic.pascal\}@centralesupelec.fr}

}

\maketitle

\begin{abstract}
This paper aims at presenting a simulative analysis of the main properties of a new $R$-estimator of shape matrices in Complex Elliptically Symmetric (CES) distributed observations. First proposed by Hallin, Oja and Paindaveine for the real-valued case and then extended to the complex field in our recent work, this $R$-estimator has the remarkable property to be, at the same time, \textit{distributionally robust} and \textit{semiparametric efficient}. Here, the efficiency of different possible configurations of this $R$-estimator are investigated by comparing the resulting Mean Square Error (MSE) with the Constrained Semiparametric Cram\'{e}r-Rao Bound (CSCRB). Moreover, its robustness to outliers is assessed and compared with the one of the celebrated Tyler's estimator.     	 
\end{abstract}


\section{Introduction}
The problem of estimating the covariance/scatter matrix from a set of observation vectors is a crucial step in many signal processing and machine learning applications, such as signal detection, clustering and distance learning. Among all the possible non-Gaussian and heavy-tailed statistical data models, the family of Complex Elliptically Symmetric (CES) distributions have been recognized to provide a general and reliable characterization of random observations in a wide range of real-word scenarios \cite{Esa}. In short, a set of $L$ i.i.d. CES distributed vectors, say $\mathbb{C}^N \ni \mb{z}_l \sim CES_N(\bs{\mu}_0, \bs{\Sigma}_0, h_0),\; l=1,\ldots,L$, is fully characterized by a location parameter $\bs{\mu}_0$, a scatter matrix $\bs{\Sigma}_0$ and a density generator $h_0:\mathbb{R}_0^+ \rightarrow \mathbb{R}^+$ that generally plays the role of a \textit{nuisance} function. In fact, inference procedures in CES distributed data usually involves the joint estimation of $\bs{\mu}_0$ and $\bs{\Sigma}_0$ in the presence of an unknown density generator $h_0$. Remarkably, this additional, \textit{infinite-dimensional}, nuisance parameter puts the CES model in the framework of \textit{semiparametric} models. Note that, due to the well-known scale ambiguity between $\bs{\Sigma}_0$ and $h_0$, only a scaled version of the scatter matrix, called \textit{shape matrix}, is identifiable. For this reason, from now on, we only consider the shape matrix $\mb{V}_{1,0} \triangleq \bs{\Sigma}_0/[\bs{\Sigma}_0]_{1,1}$ as parameter of interest instead of the unconstrained scatter matrix $\bs{\Sigma}_0$.

As recently pointed out in both the statistical \cite{Hallin_P_Annals, Hallin_Annals_Stat_2, Hallin_P_2006, PAINDAVEINE} and signal processing literature \cite{For_EUSIPCO,For_SCRB, For_SCRB_complex, Sem_eff_est_TSP}, the \textit{semiparametric} nature of the CES model allows for the derivation of semiparametric information bounds and robust inference procedures able to handle with the lack of \textit{a priori} knowledge on the actual density generator $h_0$. Common inference algorithms in CES distributed data are based on the celebrated class of robust $M$-estimators that encompasses Huber's and Tyler's estimators as special cases (see e.g. \cite{Esa}). The two main advantages of $M$-estimators of shape matrices are: \textit{i}) their performance does not drop dramatically in severe heavy-tailed scenarios and \textit{ii}) they are $\sqrt{L}$-consistent under any, possibly unknown, density generator $h_0$. On the other hand, their major drawback is their lack of (semiparametric) efficiency, as shown in \cite{For_SCRB, For_SCRB_complex}. 

In their seminal paper \cite{Hallin_Annals_Stat_2}, building upon the Le Cam's theory of Local Asymptotic Normality and on the invariance properties of rank statistics\footnote{The order statistics $Q_{L(1)}<Q_{L(2)}<\ldots<Q_{L(L)}$ of a set of (continuous) real-valued random variables $Q_1,\ldots,Q_L$ are the values of such random variables ordered in an ascending way. The rank $r_l$ of $Q_l$ is its position index in the order statistics.}, Hallin, Oja and Paindaveine have shown that it is possible to derive a shape matrix estimator able to reconcile the two dichotomic properties of \textit{robustness} and \textit{semiparametric efficiency}. This estimator, derived in \cite{Hallin_Annals_Stat_2} for Real Elliptically Symmetric (RES) distributed data, belongs to the class of rank-based, $R$-estimators. In our recent work \cite{Sem_eff_est_TSP}, a tutorial derivation of such $R$-estimator has been proposed together with its extension to CES data.

The aim of the present paper is then to validate the theoretical results about the complex $R$-estimator provided in \cite{Sem_eff_est_TSP} with a comprehensive investigation of its statistical properties. Specifically, its finite-sample performance will be analyzed in various scenarios and its semiparametric efficiency assessed by comparing its Mean Squared Error with the Constrained Semiparametric Cram\'er-Rao Bound (CSCRB) derived in \cite{For_SCRB, For_SCRB_complex}. Moreover, its robustness to \textit{outliers}, i.e. random vectors characterized by a different distribution with respect to the one of the observations, will be investigated through numerical simulations.   

\textit{Notation}: The notation used in this paper follows the one introduced in \cite{Sem_eff_est_TSP} and it is not reported here for brevity. However, for the sake of clarity, we recall the definition of some operators and special matrices that will be extensively used ahead. Specifically, $\mathrm{vec}$ indicates the standard vectorization operator that maps column-wise the entry of an $N \times N$ matrix $\mathbf{A}$ in an $N^2$-dimensional column vector $\cvec{\mb{A}}$. The operator $\ovec{\mb{A}}$ defines the $N^2-1$-dimensional vector obtained from $\cvec{\mb{A}}$ by deleting its first element, i.e. $\cvec{\mb{A}} \triangleq [a_{11},\ovec{\mb{A}}^T]^T$. A matrix $\mb{A}$ such that $[\mb{A}]_{1,1} \triangleq 1$, is indicated as $\mb{A}_1$. Let us define the following two matrices:
\be
\Pi^{\perp}_{\cvec{\mb{I}_N}}=\mb{I}_{N^2} - N^{-1}\mathrm{vec}(\mb{I}_N)\mathrm{vec}(\mb{I}_N)^T,
\ee
\be
\mb{P} \triangleq \quadre{\mb{e}_2|\mb{e}_3|\cdots| \mb{e}_{N^2}}^T,
\ee 
where $\mb{e}_i$ is the $i$-th vector of the canonical basis of $\mathbb{R}^{N^2}$. For the sake of interpretability and of consistency with the existing literature, in all our plots we show the results related to a re-normalized version of each considered estimator as:
\be\label{re_norm}
\widehat{\mb{V}}_\gamma^\varphi \triangleq N \widehat{\mb{V}}_{1,\gamma}^\varphi/\trace{\widehat{\mb{V}}_{1,\gamma}^\varphi},
\ee
where $\gamma$ and $\varphi$ indicates the estimator at hand. As performance index for the shape matrix estimators, we use:
\be
\varsigma_\gamma^\varphi \triangleq \norm{E\{\mathrm{vec}(	\widehat{\mb{V}}_\gamma^\varphi-\mb{V}_0)\mathrm{vec}(\widehat{\mb{V}}_\gamma^\varphi-\mb{V}_0)^H\}}_F,
\ee
while as performance bounds, we adopt the index \cite{For_SCRB,For_SCRB_complex}:
\be
\varepsilon_{CSCRB} \triangleq \norm{[\mathrm{CSCRB}(\bs{\Sigma}_0,h_0)]}_F.
\ee
Note that $\mb{V}_0 = N\bs{\Sigma}_0/\trace{\bs{\Sigma}_0}$ while $\bs{\Sigma}_0$ and $h_0$ represent the true scatter matrix and the true density generator. 
In all the simulations presented in this paper, we use the following common setting:
\begin{itemize}
	\item $\bs{\Sigma}_0$ is a Toeplitz Hermitian matrix whose first column is given by $[1,\rho, \ldots,\rho^{N-1}]^T$; $\rho = 0.8e^{j2\pi/5}$ and $N=8$.
	\item The zero-mean data are generated according to a complex $t$-distribution $P_{Z}(\mb{z}|\bs{\Sigma}_0, h_0)$ whose pdf is given by:
	\be
	\label{true_CES}
	p_{Z}(\mb{z}|\bs{\Sigma}_0, h_0) =|\bs{\Sigma}_0|^{-1} h_0 \left(  \mb{z}^{H}\bs{\Sigma}_0^{-1}\mb{z} \right), \; \mathrm{and}
	\ee
	\be\label{h_t}
	h_0(t) =\frac{\Gamma(\lambda+N)}{\pi^{N}\Gamma({\lambda} )}\tonde{\frac{\lambda}{\eta}}^{\lambda}\tonde{\frac{\lambda}{\eta} + t}^{-(\lambda+N)}, 
	\ee
	where $\lambda \in (1,+\infty)$ is a shape parameter that controls the tail of the distribution, while $\eta$ is a scale parameter that accounts for the data statistical power $\sigma^2$. Specifically, under the assumption of finite second order moments, we have that $\sigma^2 = \lambda/(\eta(\lambda-1))$. In our simulation, we choose $\sigma^2=4$. 
	\item The number of Monte Carlo runs is $10^6$.
\end{itemize}
It is worth underlying here that the particular choice of the complex $t$-distribution as nominal CES distribution for the observations does not represent a limitation, since, due to the semiparametric nature of the considered $R$-estimator, the findings obtained for $t$-distributed data holds true for any other CES distributions. 

\section{A robust semiparametric efficient $R$-estimator}
The aim of this Section is to recall, from an algorithmic standpoint, the $R$-estimator introduced in \cite{Hallin_Annals_Stat_2} for the RES case and in \cite{Sem_eff_est_TSP} for the CES case. We refer the reader to \cite{Hallin_Annals_Stat_2} and \cite{Sem_eff_est_TSP} for a deep theoretical investigation on its derivation and its asymptotic properties. Moreover, for the ease of exposition, in the following we assume to have a \textit{zero-mean} dataset. A procedure to handle non-zero mean data is discussed in \cite{Sem_eff_est_TSP}. For the interested reader, our Matlab and Python implementation of the algorithm can be found online at \cite{Code_R}.

Let $\mb{z}_l \sim CES_N(\mb{0}, \mb{V}_{1,0}, h_0),\; l=1,\ldots,L$ be a set of i.i.d., CES distributed, observations. The robust semiparametric efficient $R$-estimator of $\mb{V}_{1,0}$ is given by:
\be
\label{R_est}
\ovec{\widehat{\mb{V}}_{1,R}}  = \ovec{\widehat{\mb{V}}_1^\star} + L^{-1/2}\widehat{\bs{\Upsilon}}^{-1}\widetilde{\bs{\Delta}}_{\widehat{\mb{V}}_1^\star},
\ee
where $\widehat{\mb{V}}_1^\star$ is a preliminary $\sqrt{L}$-consistent estimator of $\mb{V}_{1,0}$. The matrix $\widehat{\bs{\Upsilon}}$ and the vector $\widetilde{\bs{\Delta}}_{\widehat{\mb{V}}_1^\star}$ are defined respectively as
\be
\widehat{\bs{\Upsilon}} \triangleq \hat{\alpha}\mb{L}_{\widehat{\mb{V}}_1^\star} \mb{L}_{\widehat{\mb{V}}_1^\star}^H,
\ee
\be
\widetilde{\bs{\Delta}}_{\widehat{\mb{V}}_1^\star} \triangleq L^{-1/2}\mb{L}_{{\mb{V}}_1^\star}\sum_{l=1}^{L}K\tonde{\frac{r^\star_l}{L+1}} \mathrm{vec}(\hat{\mb{u}}^\star_l(\hat{\mb{u}}^\star_l)^H)
\ee
and the scalar $\hat{\alpha}$ can be obtained as:
\be\label{com_alpha_hat}
\hat{\alpha} = \nicefrac{\norm{\widetilde{\bs{\Delta}}_{\widehat{\mb{V}}_1^\star + L^{-1/2}\mb{H}^0} - \widetilde{\bs{\Delta}}_{\widehat{\mb{V}}_1^\star}}}{ \norm{ \mb{L}_{\widehat{\mb{V}}_1^\star} \mb{L}_{\widehat{\mb{V}}_1^\star}^H\ovec{\mb{H}^0}} },
\ee
where $\mb{H}^0$ is a \virg{small perturbation}, Hermitian, matrix such that $[\mb{H}^0]_{1,1}=0$. Following \cite{Sem_eff_est_TSP}, we set $\mb{H}^0 = (\mb{G}+\mb{G}^H)/2$ where $[\mb{G}]_{i,j} \sim \mathcal{CN}(0,\upsilon^2)$, $[\mb{G}]_{1,1}=0$ and $\upsilon = 0.01$. 

The function $K:(0,1)\rightarrow \mathbb{R}^+$ is generally indicated as \textit{score function} and belongs to the set $\mathcal{K}$ of continuous, square integrable functions that can be expressed as the difference of two monotone increasing functions. 

All the other terms involved in the definition of the $R$-estimator in \eqref{R_est} are summarized as follows \cite{Sem_eff_est_TSP}:
\begin{itemize}
	\item $\hat{Q}^\star_l \triangleq \mb{z}_l^H[\widehat{\mb{V}}^\star_1]^{-1}\mb{z}_l$, 
	\item $\hat{\mb{u}}^\star_l \triangleq (\hat{Q}^\star_l)^{-1/2}[\widehat{\mb{V}}^\star_1]^{-1/2}\mb{z}_l$,
	\item $r^\star_1,\ldots,r^\star_L$ are the ranks of the (continuous) real random variables $\hat{Q}^\star_1,\ldots,\hat{Q}^\star_L$,
	\item $\mb{L}_{\widehat{\mb{V}}^\star_1} \triangleq \mb{P} \tonde{[\widehat{\mb{V}}^\star_1]^{-T/2}\otimes[\widehat{\mb{V}}^\star_1]^{-1/2}} \Pi^{\perp}_{\cvec{\mb{I}_N}}$.
\end{itemize}

In can be noted that, for a practical implementation of the $R$-estimator in \eqref{R_est}, we just need to specify two terms: the preliminary estimator $\widehat{\mb{V}}_1^\star$ and the score function $K \in \mathcal{K}$. In the following, a discussion on the most suitable choice for these two terms is provided.

\section{On the choice of the preliminary estimator $\widehat{\mb{V}}^\star_1$}
This Section is devoted to the study of the impact of the preliminary estimator $\widehat{\mb{V}}^\star_1$ on the \virg{finite-sample} performance of the $R$-estimator in \eqref{R_est}. In fact, if on one hand, any preliminary $\sqrt{L}$-consistent estimator leads to an \textit{asymptotically} semiparametric efficient $R$-estimator, on the other hand, the choice of $\pV$ may have significant impact on the \virg{finite-sample} performance of $\VR$. Here, we analyze two preliminary estimators: the Sample Covariance Matrix (SCM) and Tyler's estimator.

\subsection{The SCM as preliminary estimator}
Let $\{\mb{z}_l\}_{l=1}^L \sim CES_N(\mb{0}, \mb{V}_{1,0}, h_0)$ be a set of i.i.d. CES distributed random vectors with unknown density generator $h_0$. The SCM preliminary estimator $\widehat{\mb{V}}^\star_{1,SCM}$ is given by: 
\be
\label{SCM_1}
\widehat{\mb{V}}^\star_{1,SCM} = \frac{\hat{\bs{\Sigma}}_{SCM}}{[\hat{\bs{\Sigma}}_{SCM}]_{1,1}}, \quad \hat{\bs{\Sigma}}_{SCM} \triangleq \frac{1}{L} \sum\nolimits_{l=1}^{L} \mb{z}_l\mb{z}_l^H.
\ee 
Under the assumption of finite second order moments, $\widehat{\mb{V}}^\star_{1,SCM}$ is a $\sqrt{L}$-consistent estimator of the shape matrix $\mb{V}_{1,0}$ under any density generators, and consequently it can be used as preliminary estimator. The SCM is a very popular estimator of the covariance/shape matrix due to its low computational complexity that makes it a suitable estimator in real-time applications. However, its main drawback is in the fact that its performance rapidly decreases in non-Gaussian data. In Fig. \ref{fig:Fig1}, we report the MSE indices as function of the number $L$ of observations of $\widehat{\mb{V}}^\star_{SCM}$ in \eqref{SCM_1} and of the $R$-estimator in \eqref{R_est} that exploits $\widehat{\mb{V}}^\star_{SCM}$ as preliminary estimator. Note that both the estimators are re-normalized according to \eqref{re_norm}. As score function, we used the \textit{van der Waerden} score \cite{Sem_eff_est_TSP}: \footnote{The choice of the score function will be discussed in the next Section.} 
\be\label{K_CG}
K_{vdW}(u) \triangleq \Phi_G^{-1}(u),
\ee
where $\Phi_G^{-1}$ indicates the inverse function of the cdf of a Gamma-distributed random variable with parameters $(N,1)$. The resulting $R$-estimator will be indicated as $\widehat{\mb{V}}_{R,vdW}^{SCM}$. As we can see from Fig. \ref{fig:Fig1}, the linear \virg{one-step} correction term $L^{-1/2}\widehat{\bs{\Upsilon}}^{-1}\widetilde{\bs{\Delta}}_{\widehat{\mb{V}}_{1,SCM}^\star}$ can improve significantly the efficiency of the SCM at the price of a very small increase of computational load. Indeed, the linear \virg{one-step} correction can be evaluated in closed form, without the need of any fixed point iteration scheme required, for example, to implement an $M$-estimator. 

\subsection{Tyler's estimator as preliminary estimator}
\begin{figure}[h]
	\centering
	\includegraphics[height=5cm]{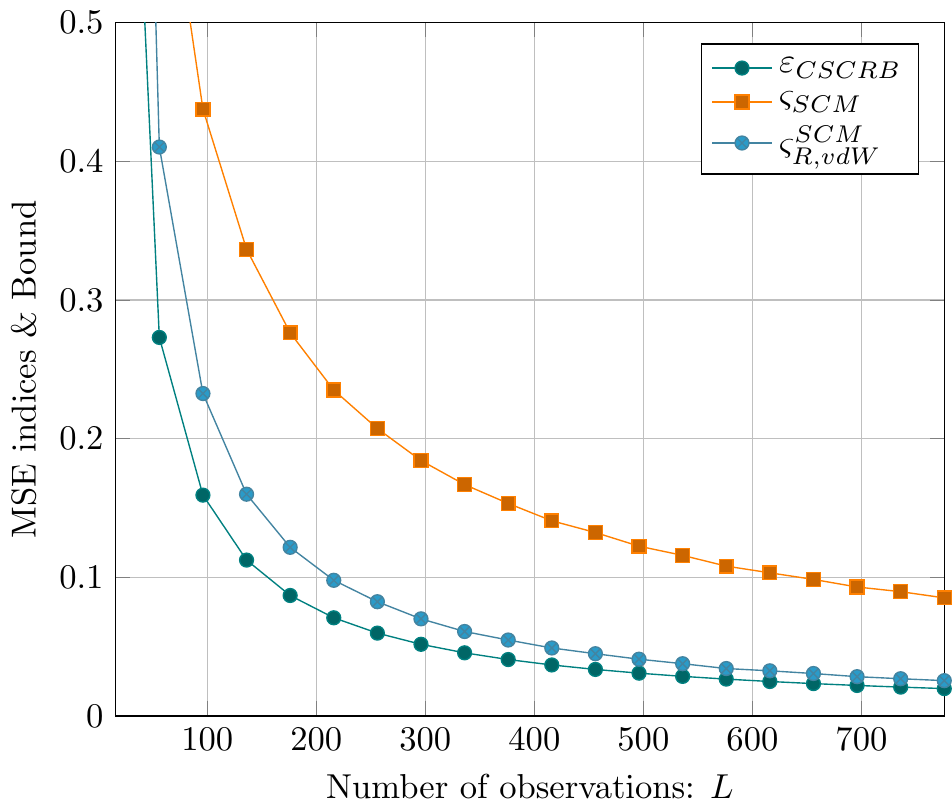}
	\caption{MSE indices and CSCRB vs $L$ ($\lambda = 2$).}
	\label{fig:Fig1}
\end{figure}

The result in Fig. \ref{fig:Fig1} has been obtained by setting a shape parameter $\lambda$ for the $t$-distributed data equal to 2. It is interesting to check the semiparametric efficiency of the $R$-estimator in \eqref{R_est} as function of $\lambda$, i.e. as function of the data \virg{heavy tailedness}, for a given number $L$ of data. Since, as said before, the SCM suffers in non-Gaussian scenarios, we may expect that the use of a robust $M$-estimator, e.g. Tyler's one, as preliminary estimator can lead to better finite-sample performance. Let us start by introducing the constrained Tyler estimator as a preliminary estimator. Tyler's estimator $\widehat{\mb{V}}_{Ty}$ can be obtained as the convergence point of the following fixed point iterative procedure: 
\be
\label{C_Tyler}
\widehat{\mb{V}}^{(k+1)}_{Ty} = \frac{N}{L}\sum_{l=1}^{L}\frac{\mb{z}_l\mb{z}_l^H}{\mb{z}_l^H[\widehat{\mb{V}}^{(k)}_{Ty}]^{-1}\mb{z}_l},
\ee
where the starting point is, e.g., $\mb{V}^{(0)}_{Ty} = \mb{I}_N$. In order to obtain a proper preliminary estimator for the $R$-estimator in \eqref{R_est}, the usual constraint on the first top-left element has to be imposed: $\widehat{\mb{V}}^{\star}_{1,Ty} = \nicefrac{\widehat{\mb{V}}_{Ty}}{[\widehat{\mb{V}}_{Ty}]_{1,1}}$.

\begin{figure}[h]
	\centering
	\includegraphics[height=5cm]{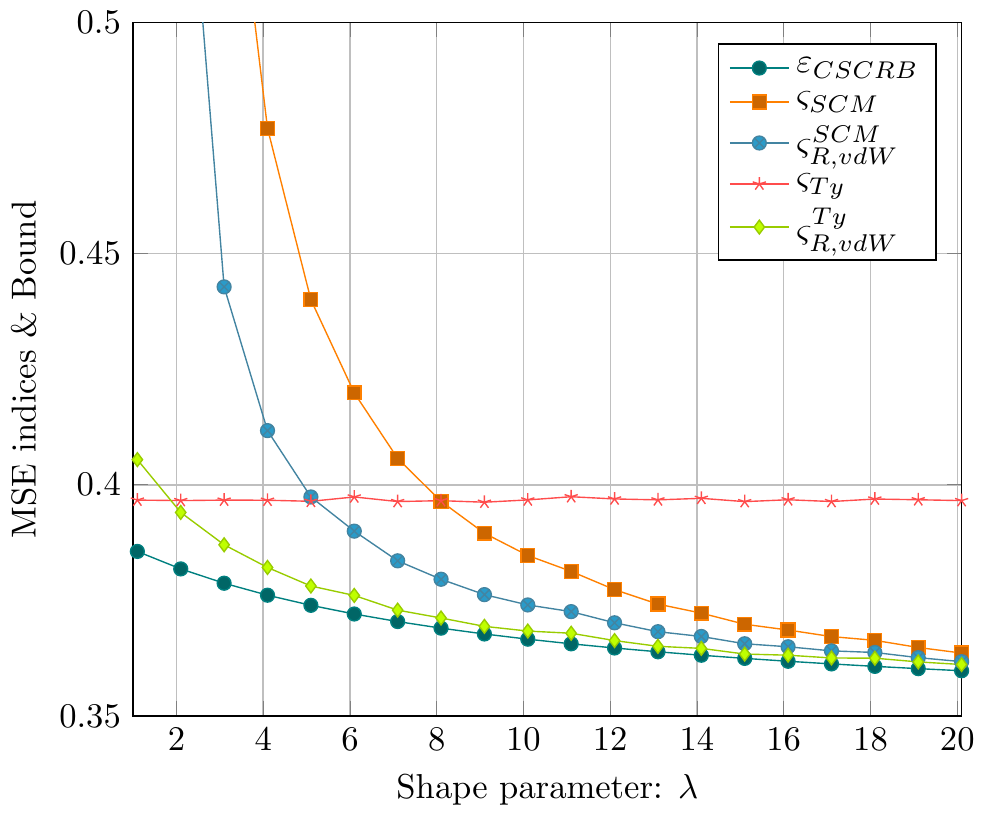}
	\caption{MSE indices and CSCRB vs $\lambda$ ($L = 5N$).}
	\label{fig:Fig2}
\end{figure}

After the re-normalization in \eqref{re_norm}, in Fig. \ref{fig:Fig2} we report the MSE indices of the SCM and Tyler's preliminary estimators, $\widehat{\mb{V}}^\star_{SCM}$ and $\widehat{\mb{V}}^{\star}_{Ty}$, together with the ones of the corresponding $R$-estimators built upon them, i.e. $\widehat{\mb{V}}_{R,vdW}^{SCM}$ and $\widehat{\mb{V}}_{R,vdW}^{Ty}$. As before, the \textit{van der Waerden} score $K_{vdW}$ in \eqref{K_CG} is used. The number of observations exploited here is equal to $L=5N$, so we are in a finite-sample regime.

As expected, $\widehat{\mb{V}}_{R,vdW}^{Ty}$ outperforms $\widehat{\mb{V}}_{R,vdW}^{SCM}$ in the presence of heavy-tailed data (small $\lambda$). This is due to the robustness property of Tyler's estimator. Clearly, the price to pay is in the computational cost needed to implement the fixed point iterations required to obtain $\widehat{\mb{V}}_{Ty}$ in \eqref{C_Tyler}. Moreover, $\widehat{\mb{V}}_{R,vdW}^{Ty}$ is an (almost) semiparametric efficient estimator for every value of $\lambda$, even in finite-sample regime. As it can be noted from \eqref{fig:Fig2}, the MSE index of $\widehat{\mb{V}}_{R,vdW}^{Ty}$ achieves the CSCRB from $\lambda >6$. Of course, in asymptotic regime, i.e. for $L \rightarrow \infty$, this semiparametric efficiency property can be achieved for smaller values of $\lambda$ as well. Another interesting point to note in Fig. \ref{fig:Fig2} is the fact that, while possessing the same distributional robustness property of the Tyler's estimator, the $R$-estimator $\widehat{\mb{V}}_{R,vdW}^{Ty}$ outperforms Tyler's one for almost all the values of $\lambda$. Remarkably, this augmented efficiency can be obtained at the negligible cost of evaluating the linear \virg{one-step} correction term $L^{-1/2}\widehat{\bs{\Upsilon}}^{-1}\widetilde{\bs{\Delta}}_{\widehat{\mb{V}}_{1,Ty}^\star}$. 

After having analyzed the impact of the choice of the preliminary estimators of the finite-sample performance of the $R$-estimator in \eqref{R_est}, in the next section we will focus on the selection of the score function $K \in \mathcal{K}$. 
 

\section{On the choice of the score function $K$}
In the context of rank statistics, the term \textit{score function} indicates a continuous scalar function $K:(0,1)\rightarrow \mathbb{R}^+$ satisfying the following two properties: \textit{i}) $K$ is square integrable and \textit{ii}) $K$ can be expressed as the difference of two monotone increasing functions \cite{Hallin_P_Annals}. A classical example of scores is the set of \textit{power functions} defined as $K_a(u)=N(a+1)u^a$ where $u\in (0,1)$ and $a \geq 0$ is a tuning parameter \cite{Hallin_PCA}. The celebrated \textit{Wilcoxon} ($a=1$) and \textit{Spearman} ($a=2$) scores belong to this set. Another way to build a score function is the one described in \cite{Hallin_P_Annals}, \cite{Hallin_Annals_Stat_2} and \cite{Sem_eff_est_TSP} where a \virg{misspecified} approach \cite{SPM} is used. We refer to \cite{Sem_eff_est_TSP} for a theoretical description of this set of scores and for a discussion on how to build them. Here we limit ourselves to cite two examples: the \textit{van der Waerden} score already introduced in \eqref{K_CG} and the $t_{\nu}$-score given by:
\be\label{CK_t}
K_{t_\nu}(u) = \frac{N(2N+\nu)F^{-1}_{2N,\nu}(u)}{\nu + 2NF^{-1}_{2N,\nu}(u)},\quad u \in (0,1),
\ee
where $F_{2N,\nu}(u)$ stands for the Fisher cdf with $2N$ and $\nu \in (0,\infty)$ degrees of freedom. Note that, from the properties if the Fisher distribution, we have: $\lim_{\nu \rightarrow \infty} K_{t_\nu}(u) = K_{vdW}(u)$.

\begin{figure}[h]
	\centering
	\includegraphics[height=5cm]{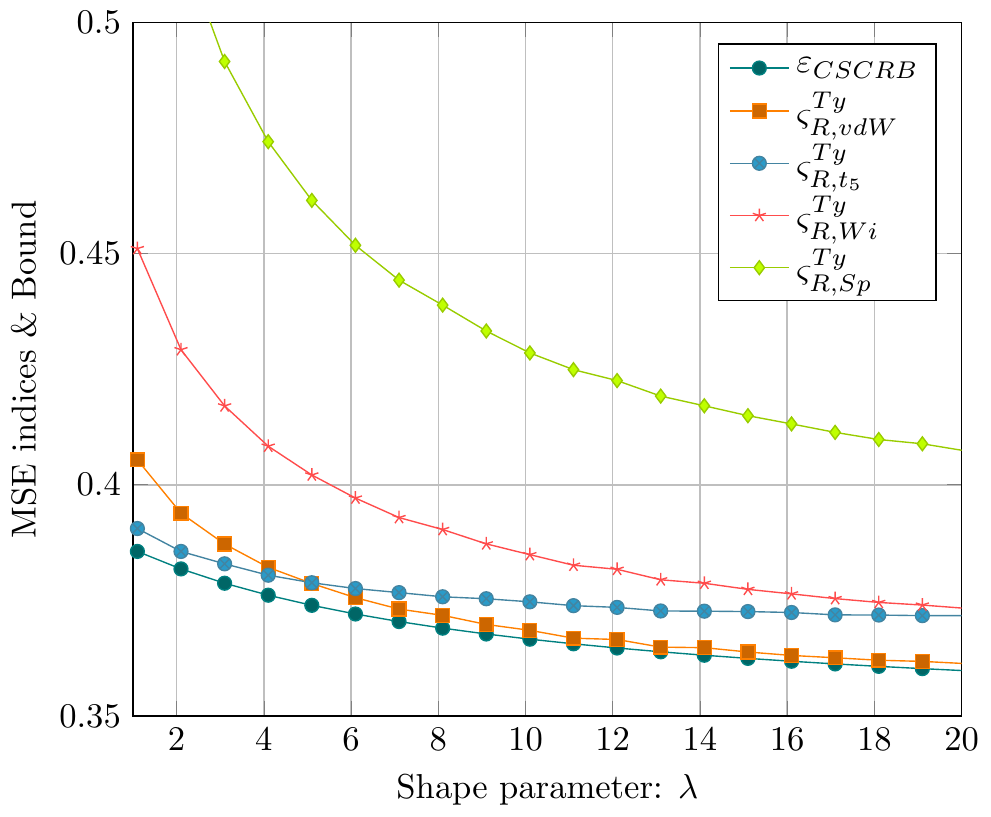}
	\caption{MSE indices and CSCRB vs $\lambda$ ($L = 5N$).}
	\label{fig:Fig3}
\end{figure}

In Fig. \ref{fig:Fig3}, after the usual re-normalization in \eqref{re_norm}, the MSE indices of four $R$-estimators exploiting as score functions the \textit{Wilcoxon} ($\widehat{\mb{V}}_{R,Wi}^{Ty}$), the \textit{Spearman} ($\widehat{\mb{V}}_{R,Sp}^{Ty}$), the $t_{\nu}$- ($\widehat{\mb{V}}_{R,\mathbb{C}t_{\nu}}^{Ty}$) and the \textit{van der Waerden} ($\widehat{\mb{V}}_{R,vdW}^{Ty}$) scores are reported and compared with the CSCRB. For the $t_{\nu}$-score, a tuning parameter $\nu=5$ has been chosen. Moreover, in all $R$-estimators, the Tyler's estimator has been used as preliminary estimator.

A visual inspection of Fig. \ref{fig:Fig3} tells us that, for $\lambda > 6$, the \textit{van der Waerden} score leads to the lowest MSE index among the other considered scores. Moreover, it can be noted that, for the power scores as the \textit{Wilcoxon} and the \textit{Spearman} ones, the resulting MSE increases as the tuning parameter $a$ increases. However, this claim should be validated by further theoretical and numerical analyses. The $t_{\nu}$-score has the best performance for highly heavy-tailed data ($1<\lambda<5$), while it provides a MSE that is between the (good) one of the \textit{van der Waerden} score and the (bad) one of the power scores. Again, $t_{\nu}$-score depends on an additional tuning parameter $\nu$ that should be carefully chosen. To conclude, we can say that the \textit{van der Waerden} score is a suitable score function since it leads to an (almost) semiparametric efficient $R$-estimator and does not depend on additional tuning parameter whose setting may result to be impractical in real-world applications.    
   
\section{Robustness with respect to outliers}
This last section is dedicated to the analysis of the robustness to \textit{outliers} of the $R$-estimator in \eqref{R_est}. An outlier is a random vector that presents a different statistical characterization with respect to the main body of the observations. In robust statistics, two main tools to quantify the robustness to outliers of an estimator are its \textit{breakdown point} and its \textit{influence function} \cite[Ch. 11 and 12]{huber_book}. The evaluation of these two fundamental tools will be left to future work. Anyway, in this Section we provide a numerical study of the robustness to \textit{outliers} of the proposed $R$-estimator by considering the Tyler's one as benchmark. More specifically, we compare the MSE indices of $\widehat{\mb{V}}_{R,vdW}^{Ty}$ and of $\widehat{\mb{V}}_{Ty}$ in two different scenarios:
\begin{enumerate}
	\item The outliers are generated as random vectors uniformly distributed on the complex unit sphere, i.e. $\mb{u} \sim \mathcal{U}(\mathbb{C}S^{N-1})$, where $\mathbb{C}S^{N-1} \triangleq \{\mb{u}\in \mathbb{C}^N|\norm{\mb{u}}=1\}$.
	\item The data are generated according to the Huber's $\varepsilon$-contamination model (see e.g. \cite[Ch. 4]{huber_book}).
\end{enumerate}

\subsection{Case 1: $\mathbb{C}S^{N-1}$-uniformly distributed outliers} 
\begin{figure}[h]
	\centering
	\includegraphics[height=5cm]{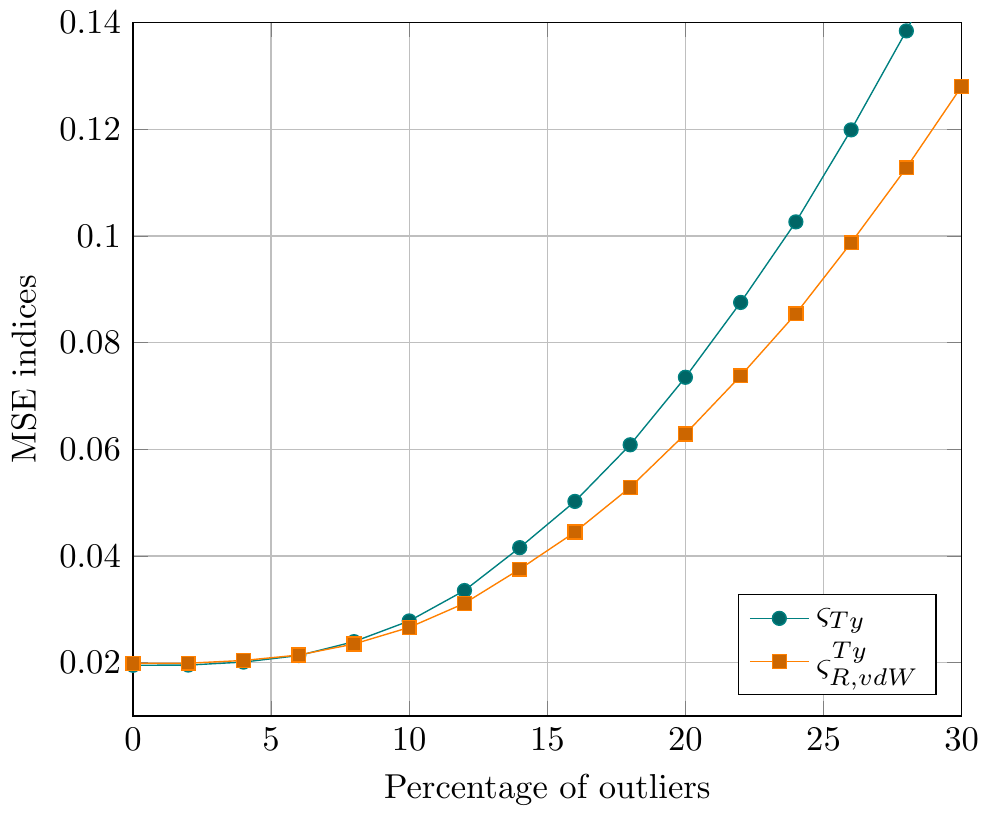}
	\caption{MSE indices vs \% of outliers ($L = 100N$, $\lambda=2$).}
	\label{fig:Fig4}
\end{figure}

In this scenario, we assume to have $L = L_p + L_o$ observations, where $L_p$ is the number of \virg{proper} observations while $L_o$ is the number of outliers. Specifically, let us assume to have a dataset 
\be
D_u \triangleq \graffe{\{\mb{z}_l\}_{l=1}^{L_p},\{\mb{u}_l\}_{l=1}^{L_o}},
\ee
where $\mb{z}_l \sim p_{Z}(\mb{z}|\bs{\Sigma}_0, h_0)$ and $h_0$ is the density generator of a $t$-distribution given in \eqref{h_t}, while $\mb{u}_l \sim \mathcal{U}(\mathbb{C}S^{N-1})$. In our numerical analysis, we use the dataset $D_u$ to estimate the shape matrix $\mb{V}_0 = N\bs{\Sigma}_0/\trace{\bs{\Sigma}_0}$ by means of the $R$-estimator in \eqref{R_est}, $\widehat{\mb{V}}_{R,vdW}^{Ty}$, and of Tyler's one in \eqref{C_Tyler}, $\widehat{\mb{V}}_{Ty}$, both re-normalized according to \eqref{re_norm}. Fig. \ref{fig:Fig4} shows the MSE indices of the two estimators as function of the percentage of outliers. As we can see, the proposed $R$-estimator presents slightly better performance than Tyler's one. Anyway, the important fact to be noted here is that the MSE of $\widehat{\mb{V}}_{R,vdW}^{Ty}$ remains stable for small percentage of outliers.   

\subsection{Case 2: $\varepsilon$-contamination model} 
The $\varepsilon$-contamination model has been firstly introduced by Huber in the context of robust hypothesis testing \cite{huber_LR}. Since then, it has been widely adopted as a suitable device to assess the robustness of both testing and estimation procedures. Let $P_{Z}(\mb{z}|\bs{\Sigma}_0, h_0)$ be the nominal data CES distribution parameterized by the scatter matrix $\bs{\Sigma}_0$ and the density generator $h_0$ and let $Q_Z(\mb{z}|\bs{\Xi}_0, l_0)$ be a \virg{contaminating} CES distribution whose scatter matrix $\bs{\Xi}_0$ and density generator $l_0$ may be different from the nominal ones. Then, the $\varepsilon$-contamination model can be expressed as the following set of distributions:
\begin{multline}\label{cont_model}
	\mathcal{F} \triangleq \left\lbrace F_Z|F_Z(\mb{z})=(1-\varepsilon)  P_{Z}(\mb{z}|\bs{\Sigma}_0, h_0)+\right. \\
	\left.  + \varepsilon Q_Z(\mb{z}|\bs{\Xi}_0, l_0), \varepsilon \in [0,1] \right\rbrace.
\end{multline}
Suppose now to have a dataset $D_c$ of $L$ i.i.d. observations sampled from a pdf $f_Z(\mb{z})$ whose related distribution $F_Z(\mb{z})$ belongs to $\mathcal{F}$ in \eqref{cont_model}, i.e.:
\be\label{Z_cont}
D_c=\{\mb{z}_l\}_{l=1}^L, \; \mb{z}_l \sim f_Z(\mb{z}).
\ee
This implies that, with probability $(1-\varepsilon)$, the $l^\mathrm{th}$ observation vector $\mb{z}_l$ has distribution $P_{Z}(\mb{z}|\bs{\Sigma}_0, h_0)$ (i.e. it is a valuable observation) while, with probability $\varepsilon$, $\mb{z}_l$ has distribution $Q_Z(\mb{z}|\bs{\Xi}_0, l_0)$ (i.e. it is an outlier).

In our simulations, we assume as nominal distribution $P_{Z}(\mb{z}|\bs{\Sigma}_0, h_0)$ the complex $t$-distribution whose relevant pdf is given in \eqref{true_CES}. As contaminating CES distribution $Q_Z(\mb{z}|\bs{\Xi}_0, l_0)$ we adopt a Generalized Gaussian (GG) distribution whose density generator can be expressed as \cite{Esa}:   
\be\label{l_t}
l_0(t) =\frac{s\Gamma(N)b^{-N/s}}{\pi^{N}\Gamma(N/s )}\exp\tonde{-t^s/b}, 
\ee
As parameters characterizing the GG distribution, we choose $\bs{\Xi}_0 \triangleq \sigma^2\mb{I}_N$ as scatter matrix, $s=0.1$ as shape parameter while $b$ is set in order to provide for the outliers the same statistical power $\sigma^2$ of the $t$-distributed data.  

\begin{figure}[h]
	\centering
	\includegraphics[height=5cm]{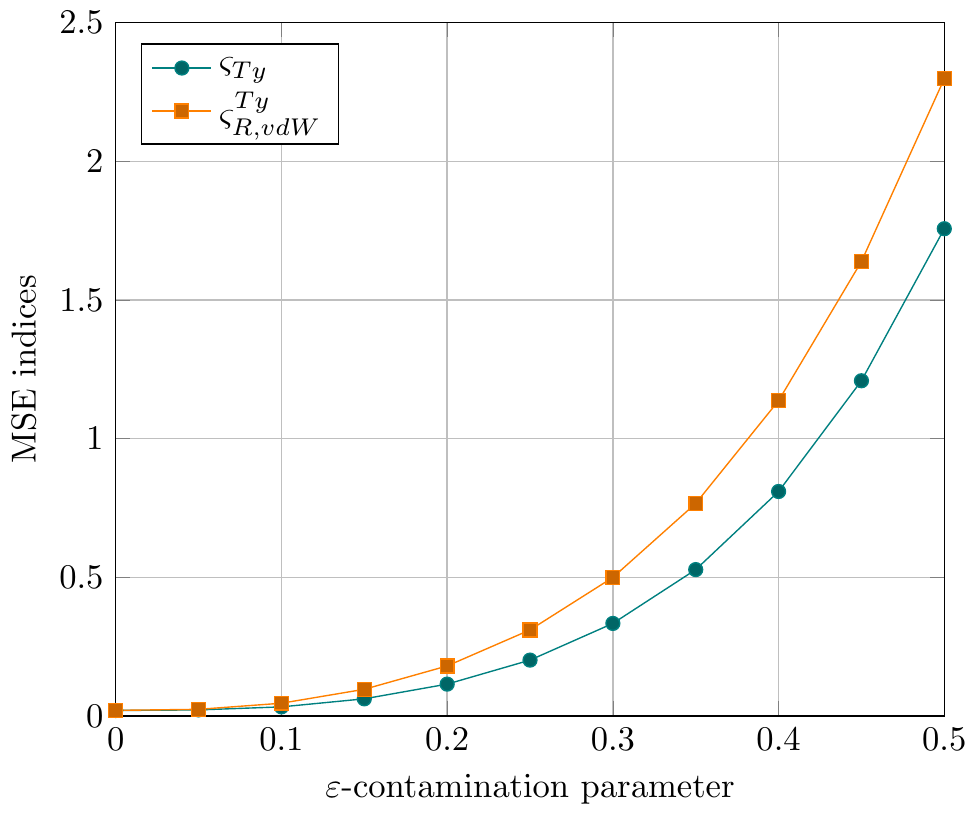}
	\caption{MSE indices vs $\varepsilon$ ($L = 100N$, $\lambda = 2$).}
	\label{fig:Fig5}
\end{figure}

As for the Case 1 previously discussed, we estimate the shape matrix $\mb{V}_0 = N\bs{\Sigma}_0/\trace{\bs{\Sigma}_0}$ from the contaminated dataset $D_c$ in \eqref{Z_cont} by means of $\widehat{\mb{V}}_{R,vdW}^{Ty}$ and of $\widehat{\mb{V}}_{Ty}$, both re-normalized according to \eqref{re_norm}. The resulting MSE indices are reported in Fig. \ref{fig:Fig5} as function of the $\varepsilon$-contamination parameter. It can be noted that, even if the Tyler's estimator have slightly better performance than the $R$-estimator, the MSE of $\widehat{\mb{V}}_{R,vdW}^{Ty}$ does not increase dramatically as $\varepsilon$ increases and it remains close to the one relative to $\widehat{\mb{V}}_{Ty}$.  

\section{Concluding remarks}
In the first part of this paper, the \virg{finite-sample} performance of the proposed $R$-estimator have been analyzed in different configurations and compared with the relevant CSCRB. The proposed numerical investigation showed that the setting involving Tyler's estimator as preliminary estimator and the \textit{van der Waerden} score as score function leads to the lowest MSE index among all the other considered configurations and for almost all values of $\lambda$. Furthermore, the robustness to outliers of the $R$-estimator has been assessed by using the one of Tyler's estimator as benchmark. The proposed numerical study revealed that the $R$-estimator is approximately as robust to outliers as Tyler's one. Needless to say, the numerical analysis provided in this paper is just a first attempt but a theoretical characterization of the robustness in terms of breakdown point and influence function is necessary and it is left to future works.

\bibliographystyle{IEEEtran}
\bibliography{ref_semipar_eff_estim}

\end{document}